# An Analytic and Experimental Treatment of Fiber Optic Chemical Sensing: Results on Evanescent Wave Spectrometry

Joseph Plumitallo[1], Jin Ho Kim[1], Silverio Johnson[1], Do-Joong Lee[1], Stephen Giardini[2], Sean Dinneen[2], Richard Osgood III[2], Jimmy Xu[1]

[1] Brown University (United States)
[2] US Army Combat Capabilities Development Command Soldier Center (United States).

## Abstract

Standoff chemical sensing of chemicals and toxic compounds is very desirable for actively monitoring residential, commercial, and industrial spaces. Analytes such as toxic industrial compounds (TIC), chemical warfare agents (CWAs), and environmental pollutants (EPs) can pose serious threats to civilians, warfighters, and the environment. We report on highly sensitive, robust, and inexpensive distributed standoff fiber optic chemical sensors (FOCSs). FOCSs are activated by analytes reaching a sensing region and inducing a change in its optical properties. The information of a spectral response is transmitted by a shift in transmission spectra probed by evanescent waves, corresponding to the detection of an analyte. However, there are many factors that play a role in the implementation and optimization of such fiber sensors. We find it beneficial to construct the theory of general optochemical waveguide sensors to guide the analysis. We apply it to a specific fiber platfom that is robust, broadband, multi-target, scalable, and compatible to texitle fibers and fabrics – a focus of this work. The good agreement between the predicted and measured spectra of FOCSs exposed to Ammonia ($NH_3$) lends support to the analysis. The experiment yields a 1ppm sensitivity, representing the current best in class. The use of $\mathrm{NH_3}$ in the test serves as a representative for a larger set of amine-based TICs, CWAs, EPs, and other chemical species, targeted for wide-area distributed coverage and textile-compatible sensors.

## Introduction

Point and standoff sensors are an established and growing part of our daily lives. Their advancement into distributed sensors is important to cover more sensing area to reach locations that may be difficult for human observers, and for which point detectors, even a network of point detectors, is insufficient. For example, in a home fire alarm network – an example of a network of point sensors – if one alarm fails to detect smoke or CO emissions, then the entire network fails to detect the onset of a fire. In contrast, a distributed sensor takes advantage of its length to increase sensitivity to small signals, i.e., low-concentration analyte. As long as the noise doesn't scale in the same way as the optical absorption along the fiber's length due to chemical absorption in the cladding, it is reasonable to expect that longer fibers will be sensitive to the lower concentrations for the same minimum detectable dose, assuming absorption all along the fiber's length. In addition to wider coverage and detecting smaller concentrations than that of a point detector, a distributed detector like a FOCS is capable, in principle, of making spatially resolved measurements.

Sensors also need to be more rugged, selective, flexible, inexpensive, and reversible. Sensors collect information to better understand any given system, e.g., for advanced medical diagnostics, atmospheric monitoring, automotive diagnostics, gas leaks, and so on [7][32][36][39] [40]. Sensors can be classified into three major categories: physical sensors, chemical sensors, and biological sensors [5]. While fiber optical chemical sensing involves important chemistry, the sensing performance also depends on physics.

Chemical sensors are a large subcategory of sensors that play a critical role in everyday state of affairs, from protecting homeowners from carbon monoxide (CO) poisoning to monitoring greenhouse gases such as methane ($\mathrm{CH_4}$) [27]. Gaseous state chemical sensing methods can be many and implemented in various platforms [19]. We focus on fiber optic chemical sensors, which fall under the category of chemical optical waveguide sensors, or optochemical waveguide sensors.

Optical fibers themselves are typically placed in the context of telecommunication technologies, but it is interesting to note that initial efforts were strongly focused on sensing devices rather than communications [10]. The foundations of fiber optics can be traced as early as the mid-1800s, although there was no significant progress in practical applications until pioneering work done by nobel laureate Charles K. Kao [12][33]. Since Kao's work, there has been aggressive and fruitful advancements in fiber optic technologies, ranging far and wide from biosensors to photovoltaic power transmission [10] [30] [34] [36].

We report on polymer FOCSs coated with a cylindrical section of sensing dye whose colorimetric properties are measured by evanescent field spectrometry. The sensing fibers are constructed with a commercially available multi-mode polymer core wrapped in step-indexed cladding. They are robust, flexible, broadband, low-loss, scalable (to hundreds of meters), and compatible to textile fibers and fabrics which is of particular interest to us at present. To enable effective sensing, a small cylindrical section of the fiber's cladding is chemically etched and the exposed core is coated with a sensing dye. This dye region is the functional sensing region which allows the detection of target analytes, whether in the form of TICs, CWAs, EPs, or other toxic (even innocuous) compounds.

A good FOCS should have good selectivity, robustness, reversibility and the ability to detect low concentrations of target analytes. A reversible sensor is one that is restorable to its initial state after a sensing event [14] – desirable, for example, in continuous standoff sensing and reporting. Other desirable characteristics include low cost, portability, safety, and scalability. Each gas sensor type usually bears some tradeoff between advantages and disadvantages under different circumstances, ultimately falling into its own niche. For example, metal-oxide gas sensors are a good option under high temperature environments, but they suffer from poor selectivity. On the other hand, conducting polymer gas sensors generally have good selectivity, but they suffer from poor stability [18]. In this spirit, we will explore the set of advantages and disadvantages of FOCSs that are robust, sensitive, selective, reversible, portable, scalable, inexpensive, safe, and in particular, compatible to textile-fibers and fabrics. Preferrably FOCS should also exhibit tolerance to humidity and interferents [25].

We first introduce a general waveguide theory to set up a theoretical framework within which optical fibers fall into. Thereafter, we constrain the more general waveguide theory to dielectric cylindrical waveguides (optical fibers), and later our specific construction. Smaller diameter, so-called single-mode fibers, are analytically tractable, as they only admit one mode of light propagation. On the other hand, larger diameter, so-called multi-mode fibers, are much more analytically cumbersome, usually requiring numerical simulations or a challenging full-wave analysis. However, we prefer multi-mode fibers for FOCSs because they afford the low-loss broadband transmission that is more suited for chemical sensing of multiple targets. In order to help the analysis guide our design, we analyse the modes that exist in a multi-mode fiber and employ the weakly guiding approximation to ultimately simplify the effort necessary to examinethe spectral response of our broadband sensing fibers. Finally, we compare signiture features of our experimental results with those predicted by the corresponding model for validation and design guidance of further improvement.

## Theory of Waveguides

Before modelling the propagation of light through an optical fiber, we will first lay out a more general theory of light propagation through a general waveguide structure. A waveguide structure, in general, is one that guides waves (of any nature) along a preferred direction such that energy loss is minimized, such as optical fibers. So, it is therefore desirable to first generate a general theory of waveguides and then impose a set of constraints that correspond to cylindrical dielectric waveguide structures. This will allow us to finally produce a model for the propagation of light through an optical fiber [9] [13] [23] [29]. We first assume the waveguide is free of local currents and fields. Then, Maxwell's equations take a familiar form:

$$\nabla \cdot \vec{E} = 0 \qquad \nabla \cdot \vec{B} = 0$$
$$\nabla \times \vec{E} + \frac{\partial \vec{B}}{\partial t} = 0 \qquad \nabla \times \vec{H} - \frac{\partial \vec{D}}{\partial t} = 0 \qquad (1)$$

Assuming time dependence of the form $e^{i\omega t}$, and linear, isotropic, media ($\vec{D} = \epsilon \vec{E}$ and $\vec{B} = \mu \vec{H}$), taking the curl of the Ampere-Maxwell and Farady equations gives rise to the familiar Helmoltz wave equations:

$$(\nabla^2 + \mu\epsilon\omega^2)\vec{E} = 0 \qquad (3)$$

$$(\nabla^2 + \mu\epsilon\omega^2)\vec{B} = 0 \qquad (4)$$

Let's take our preferred direction along the z-axis for the $\vec{E}(x,y,z,t)$ and $\vec{B}(x,y,z,t)$ fields as follows:

$$(\vec{B})\vec{E}(x,y,z,t) = (\vec{B})\vec{E}(x,y)e^{\pm ikz - i\omega t} \tag{5}$$

It is convenient to separate operators and fields into components that are tranverse and parallel to the z-axis:

$$\nabla_t^2 \equiv \nabla^2 - \frac{\partial^2}{\partial z^2} \tag{6}$$

$$\vec{E} = \vec{E}_z + \vec{E}_t \quad (\text{similarly for } \vec{B})$$

$$\text{where } \vec{E}_z \equiv E_z \hat{z} \text{ and } \vec{E}_t = (\hat{z} \times \vec{E}) \times \hat{z} \tag{7}$$

The Helmholtz equations paired with Dirichlet and Neumman boundary conditions give us the standard transverse electromagnetic (TEM), transverse electric (TE), and transverse magnetic (TM) modes. TEM waves are conveniently derivable from a potential such that they are plane waves with wavenumber $k = \pm\sqrt{\mu\epsilon\omega^2} = \pm v\omega$, and one can recall that TEM waves are unphysical inside of conducting waveguides. Treating the Helmholtz equation as an eigenvalue problem with a wave function, call it $\psi$, the number of modes then become finite:

$$(\nabla_t^2 + (\mu\epsilon\omega^2 - k^2))\psi = 0 \tag{8}$$

which is solved for chosen neumman or dirichlet boundary conditions (TE or TM). A general solution to the Helmholtz eigenvalue problem will give rise to a respective orthogonal eigenspectrum. For notational convenience, take the eigenvalues $(\mu\epsilon\omega^2 - k^2) \equiv \gamma_\lambda^2$, and the corresponding solutions $\psi_\lambda$, for $\lambda \in \mathbb{N}$. Each wave solution $\psi_\lambda$ is called a mode. Now, from the Helmholtz equation, we also see the the eigenvalues must be positive, otherwise $\psi_\lambda$ would exhibit exponentially decaying features – clearly not wave modes. Thus, $\mu\epsilon\omega^2 \geq k^2$. Thus, we define the cutoff frequency, at which modes cease to exist:

$$\omega_\lambda = \frac{\gamma_\lambda}{\sqrt{\mu\epsilon}} \tag{9}$$

As optical fibers are waveguides, we now discuss the results of reducing general waveguide theory to that of optical fibers [13]. A typical optical fiber consists of a cylindrical dielectric core material with refractive index $n_1$, wrapped in some cladding material with a refractive index of $n_2$ (fig. 1). This is called a step-index fiber, in contrast to a graded-index fiber, which has a radial gradient of indices of refraction. We will not concern ourselves with graded-index fibers (mostly used in telecom). The most important metric for choosing a cladding material is its refractive index. Snell's law of refraction shows us that the relative indices of refraction and the angle of light incidence (with respect to the surface norm), together, largely determine whether or not light will continue to propagate through the waveguide:

$$\left[\frac{\sin(i)}{\sin(r)} = \frac{n_2}{n_1}\right]_{r=\frac{\pi}{2}, i=i_0} \quad i_0 = \sin^{-1}\left(\frac{n_2}{n_1}\right) \tag{10}$$

Here, following the convention and derivation of Jackson [13], $i$ is the angle of incidence (defined from the normal to the interface which lies along the radial axis for an optical fiber), $r$ is the angle of refraction, $n_1$ is the refractive index of the fiber core, and $n_2$ is the refractive index of the fiber cladding.

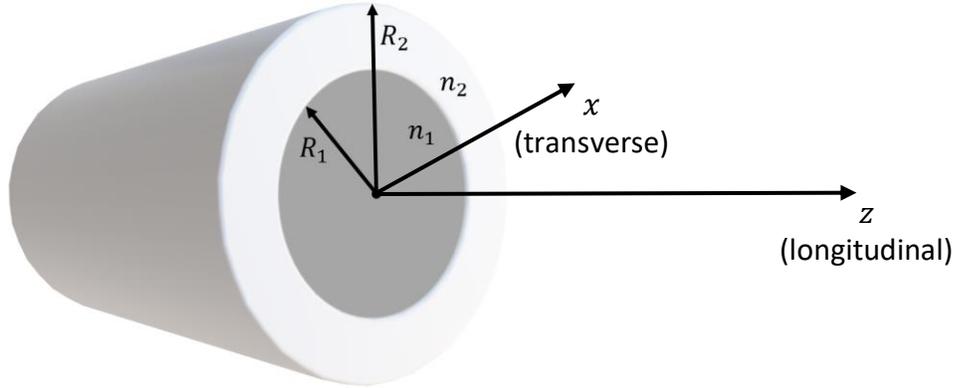

Figure 1: A typical optical fiber configuration with a core material of radius $R_1$ (for polymer multi-mode fibers, usually $\sim 500 - 1000 \mu m$) and real refractive index $n_1$, and a cladding material of radius $R_2$ (usually $\sim 550 - 1050 \mu m$) and real refractive index $n_2 < n_1$

The result of (10) is that of total internal reflection, and occurs at the angle of incidence $i_0$. For $i > i_0$, total internal reflection implies the angle of refraction is complex (since $\sin(r) > 1$) with a purely imaginary cosine:

$$\cos(r) = i\sqrt{\left(\frac{\sin(i)}{\sin(i_0)}\right)^2 - 1} \tag{11}$$

Thus, the propagation factor for the traveling wave goes as follows (note a different reference frame than in Ref. [13]):

$$e^{i\vec{k}\prime\cdot\vec{r}} = e^{ik\prime(z\sin(r)+x\cos(r))} = e^{-k\prime x\sqrt{\left(\frac{\sin(i)}{\sin(i_0)}\right)^2-1}} e^{ik\prime z\left(\frac{\sin(i)}{\sin(i_0)}\right)} \tag{12}$$

Here, we see that when $i > i_0$, the refracted wave travels along the interface, and is attenuated in the x-direction by the decaying term $exp\left\{-k'\left[\left(\frac{\sin(i)}{\sin(i_0)}\right)^2 - 1\right]^{\frac{1}{2}}\right\}$. Power is only transmitted along the z-direction; e.g., along the core of the optical fiber [13], since there is no power flow into the cladding [13].

Other parameters of optical fibers are the number and behavior of modes within a given fiber. Since we are more interested in relatively thick fibers for easy in- and out-coupling of broad-band lights in field and portable sensings, single-mode telecom fibers are excluded. To characterize such optical fibers, it is useful to understand approximately how many modes can propagate within a fiber. This can be done using phase space arguments. The number density in this phase space goes as follows:

$$N = r^2 k^2 \int_0^{\theta_{max}} \theta d\theta \approx \frac{1}{2}(kr\sqrt{2\delta})^2 \quad (13)$$

where $\theta_{max}$ is the cutoff angle and $\delta \equiv \frac{1}{2}\left(1 - \left(\frac{n_2}{n_1}\right)^2\right)$, and when $\delta \ll 1$, we can invoke the so-called weakly guiding approximation ($kr\sqrt{2\delta}$ is sometimes referred to as the "V parameter"). For example, if we take a radius of $500\mu m$, $n_1 \approx 1.5$, $n_2 \approx 1.49$, and $\lambda = .6\mu m$, we get a number of modes on the order of $10^5$. This may appear to be rather large, but it is beneficial to the development of broad-band chemical sensing platforms and allows for multi-target identification by permitting analyses of complex spectral bands of chemical reactions over full ranges of functional wavelengths. There are two classes of optical fiber modes: meridonial rays and skew rays. Meridonial rays are those wich pass through the cylinders axis; i.e., nonvanishing intensity along the longitudinal axis. Skew rays, on the other hand, are those that do not pass through the cylinders axis, traveling along a helical trajectory; i.e., vanishing intensity along the longitudinal axis (fig. 2).

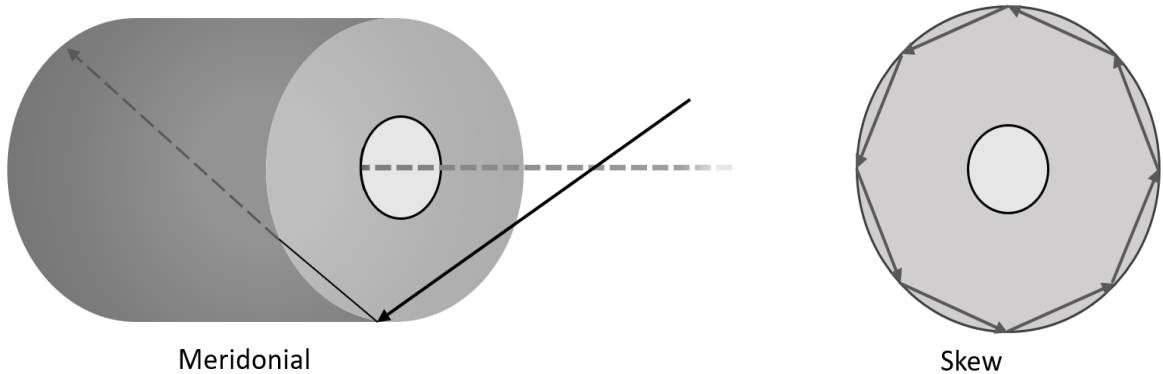

Meridonial  Skew

Figure 2: A visualization of meridonial rays versus skew rays. The cladding material is omitted here, and pictured is just the core structure. Here, we see the two dimensional projection of the helical path of a skew ray (right). On the other hand, the schematic of the meridonial ray shows the ray passing through the longitudinal axis (left).

These rays assume plane wave propagation. As we will see, the modes are not quite plane waves, but for our purposes, this is a satisfactory approximation in the weakly guiding regime. Under the boundary conditions of a cylindrical dielectric waveguide, the Helmholtz equations give us $\vec{E}$ and $\vec{B}$ field coupled transverse components:

$$\nabla_t^2 H_z + \gamma^2 H_z - \left(\frac{\omega}{\gamma c}\right)^2 (\nabla_t n^2) \cdot \nabla_t H_z = -\frac{\omega k_z \epsilon_0}{\gamma^2} \hat{z} \cdot [\nabla_t n^2 \times \nabla_t E_z] \quad (14)$$

$$\nabla_t^2 E_z + \gamma^2 E_z - \left(\frac{k_z}{\gamma n}\right)^2 (\nabla_t n^2) \cdot \nabla_t E_z = \frac{\omega k_z \mu_0}{\gamma^2 n^2} \hat{z} \cdot [\nabla_t n^2 \times \nabla_t H_z] \qquad (15)$$

As the fields are coupled, we find the hybrid electric (HE) and hybrid magnetic (HM) modes with solutions as bessel functions respectively inside of the fiber:

$$E_z = A_e * J_m(\gamma \rho) e^{im\phi}, \qquad \rho < r \qquad (16)$$

$$H_z = A_h * J_m(\gamma \rho) e^{im\phi}, \qquad \rho < r \qquad (17)$$

for $\gamma \equiv n_1^2 \frac{\omega^2}{c^2} - k_z^2$, $\phi$ as the azimuthal angle, and $\rho$ as some interior radial distance. If we satisfy the weakly guiding approximation, then one can show that the bessel modes $E_t$ and $H_t$ are much larger in magnitude than the longitudinal bessel (and modified bessel) modes, $E_l$ and $H_l$. Thus, we have nearly recovered TEM (plane-like) waves that are virtually linearly polarized, and vary transversely as $J_m(\gamma \rho)$ – these are our LP modes [13] [29]. Thanks to this simplification, this is how we shall proceed when characterizing these FOCSs.

## Fiber Optic Chemical Sensing (FOCS)

We can next extend the framework of the analysis to model desirable properties in designing the FOCSs [10] [30]. Among the many configurations of FOCSs, there is a general structure that underpins most FOCS devices; that is, there is some light source transmitted through the fiber to some sensing region, which is then transmitted to a detecting region. For broad-band multi-target chemical sensing, we couple a halogen bulb light source (Thor Labs OSL2 broadband Halogen fiber optic illuminator) to a multi-mode step-index optical fiber with a polymethylmethacrylate (PMMA) polymer core. This fiber is mechanically sturdy, optically low-loss, and compatible to fabrics and portable applictions. The light that exits the fiber is coupled to a (Hamamatsu C12880MA) microspectrometer. Our sensing region consists of a small section ($\approx 20cm$) of the fiber where a solution of Bromothymol Blue (BTB) and Ethyl Cellulose (EC) is bound to the core of the fiber, filling the space where cladding was chemically etched. The chemical etching and binding process begins with a cylindrical section of the fiber ($\approx 20cm$) submerged in pure acetone for 10 seconds, and thereafter immediately submerged in ethanol for 10 seconds. Three iterations of this back-and-forth between solutions are done for a total of 60 seconds. The process ends by pipetting the BTB-EC solution onto the exposed fiber core and allowing it to cure overnight (fig. 3).

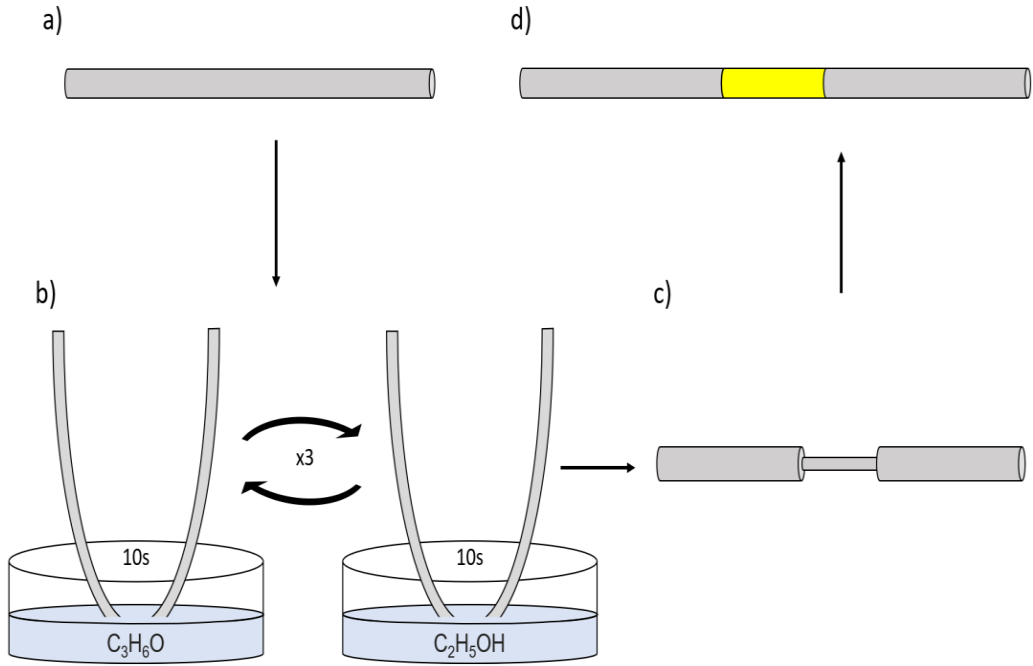

Figure 3: This is a schematic of the fabrication process for the fiber component of the FOCS. It is a four step process that begins with a commercially available PMMA core multi-mode fiber (a). From this fiber, an approximately 20cm cylindrical section is chemically etched by alternatingly submerging that section in acetone and ethanaol for 10 seconds each, over three iterations, for a total of 60 seconds (b-c). A BTB-EC solution is then pipetted onto the exposed core of the fiber (d), and it is left to cure overnight.

Once the fiber is prepared, one end can be coupled to a light source and the other end can be coupled to a spectrometer for sensing applications. The process of detection is as follows (fig. 4):

1. Light from a source enters the fiber, traveling with small attenuation until it reaches the sensing region.

2. Once the light reaches the sensing region, it interacts with the sensing dye and shifts its transmission spectra accordingly. Either a target analyte has reached the sensing region, or it has not.

   (a) If there are no analytes in the sensing dye region, then the chemical composition of the BTB-EC solution (and therefore real and complex refractive indices) remain constant, and our transmission spectra has some footprint corresponding to what we will call an initial state.

   (b) If instead analytes have interacted with the sensing dye region, then the sensing dye will undergo a chemical compositional (pH) change, leading to a physical coloration response, and therefore the transmission spectra of the light will shift from its initial state into a new state.

3. Lastly, we observe the spectrometer to monitor any shifts in transmission spectra. A clear shift from the initial state into our expected reaction state is a good indication of the presence of a target analyte.

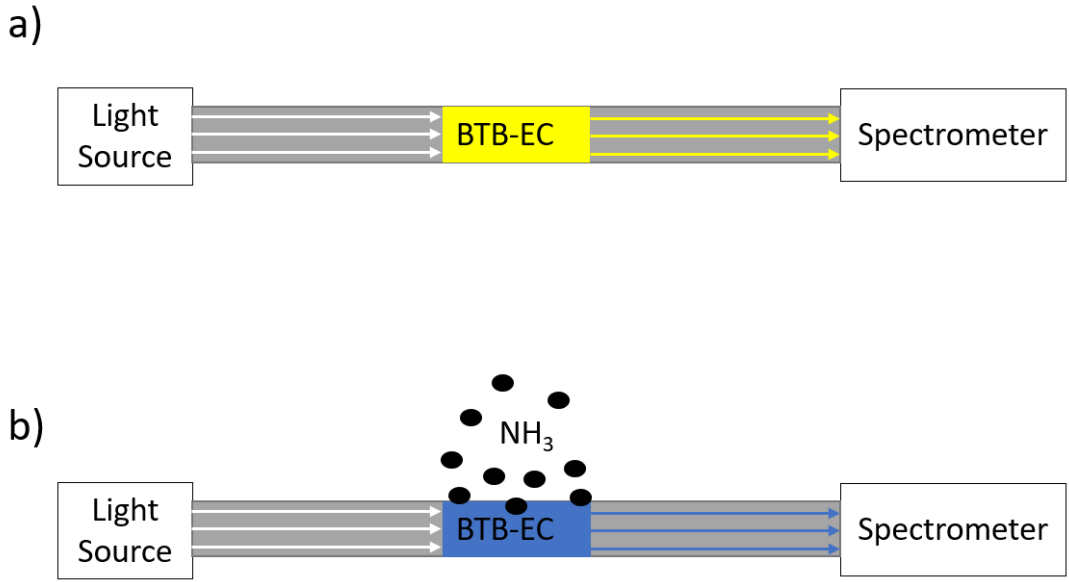

Figure 4: A schematic of the FOCS's detection process (a) before analytes reach the sensing dye, and (b) after analytes reach the sensing dye. The shift in transmission spectra detected by the spectrometer is physically realized by a physical coloration response via the BTB-EC coating.

Through this process, and the description of this setup, we can begin to see the desirable characteristics of this sort of FOCS: safety, as there are no harmful or toxic components in the fiber or its sensitizing process; low cost, as PMMA fibers, dye solutions, and microspectrometers are relatively inexpensive; robust, as polymer fibers are flexible and strong, and the collective components of this setup are seen to be tolerant of many chemicals, fluids, electromagnetic radiation, temperature fluctuations, pressure fluctuations, etcetera; portability, as the whole setup is lightweight; responsive, as the information travels at the speed of light, and chemical diffusion is usually on the order of seconds; and scalable, as the fibers can range in length from millimeters to hundreds of meters [25]. In particular, they are compatible to texitle-fibers and fabrics which can be viewed as a new space of opportunities for chemical sensing. Selectivity and reversibility will also be seen to be relatively strong, and will be discussed in more detail.

## Experimental Results

The analysis on optical fiber theory showed that the weakly guided approximation to a dielectric multi-mode cylindrical fiber gives rise to near plane waves that vary transversely as $J_\mathrm{m}(\gamma\rho)$ – a good approximation for the setup that we are studying. We have collected data (from ellipsometry experiments) for the real and complex components of the refractive index of a BTB-based sensing dye solution applied to an optical fiber before and after it was exposed to trace ammonia $\mathrm{NH_3}$ gas. With the plane wave approximation, one can then perform an analysis on attenuation over a representative sample of wavelengths to predict the shift in transmission spectra as follows:

$$\vec{E} \cong \vec{E}_0 e^{i\vec{k}\cdot\vec{r}} = \vec{E}_0 e^{ikz} = \vec{E}_0 e^{i\frac{\omega}{c}nz} = \vec{E}_0 e^{i\frac{\omega}{c}n_r z} e^{-\frac{\omega}{c}n_i z} \qquad (18)$$

With this information, we can calculate the attenuation of intensity as a function of the active set of wavelengths we should expect based on the emission spectra of the light source and the sensing dye refractive indices.

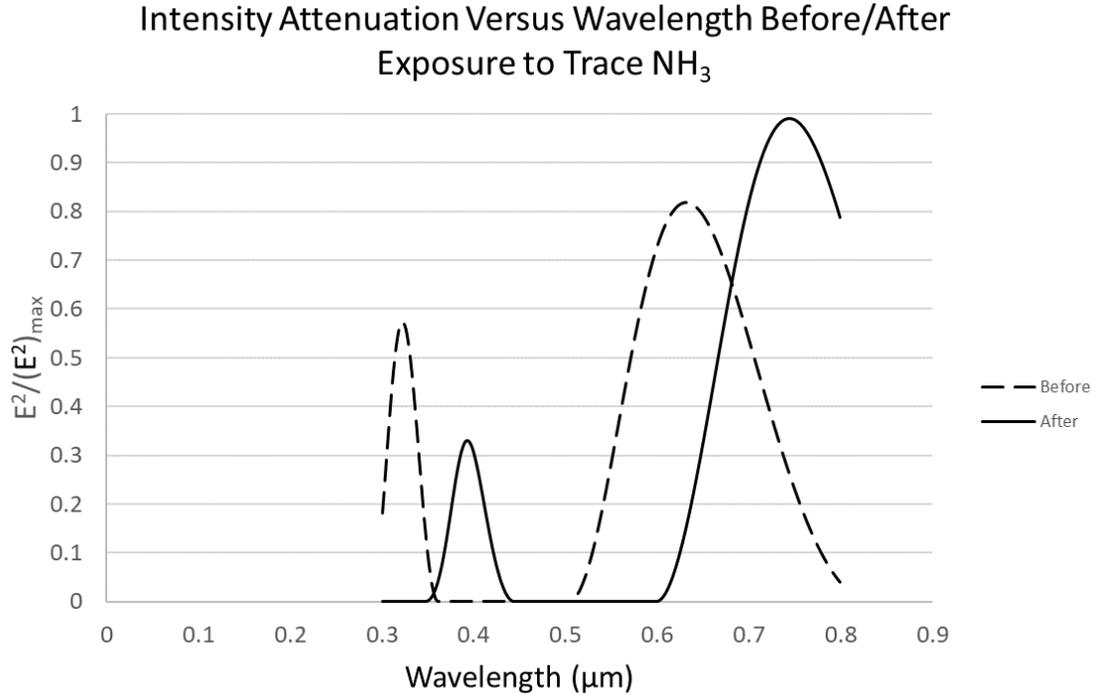

Figure 5: Graph showing the calculated strength of intensity attenuation as a function of wavelength. Before exposure to trace $NH_3$ (dotted curve), we expect strong attenuation in the blue region, giving rise to an orange/reddish color. After exposure to trace $NH_3$ (solid curve), we expect strong attenuation in the green and yellow regions, giving rise to a bluish/purplish color, as indicated by the dye's name, bromothymol blue.

We see (fig. 5) that before exposure to trace $NH_3$ gas, there is strong absorption in the ~350-500nm region, and little absorption in the ~300-350nm and ~550-700nm regions; thus, we should initially observe an orange/reddish color, and expect spectral intensity peaks (mindful of the emission spectra of the light source) in the ~300-350nm and ~550-700nm regions. Additionally, after exposure to trace $NH_3$ gas, there is strong absorption in the ~450-600nm region, and little absorption in the ~350-450nm and ~700-800nm region; thus, we should observe a bluish/purplish color (hence bromothymol blue) after a detection event, and expect a spectral intensity peak in the ~550-700nm region. This analysis supported by experiment (fig. 6).

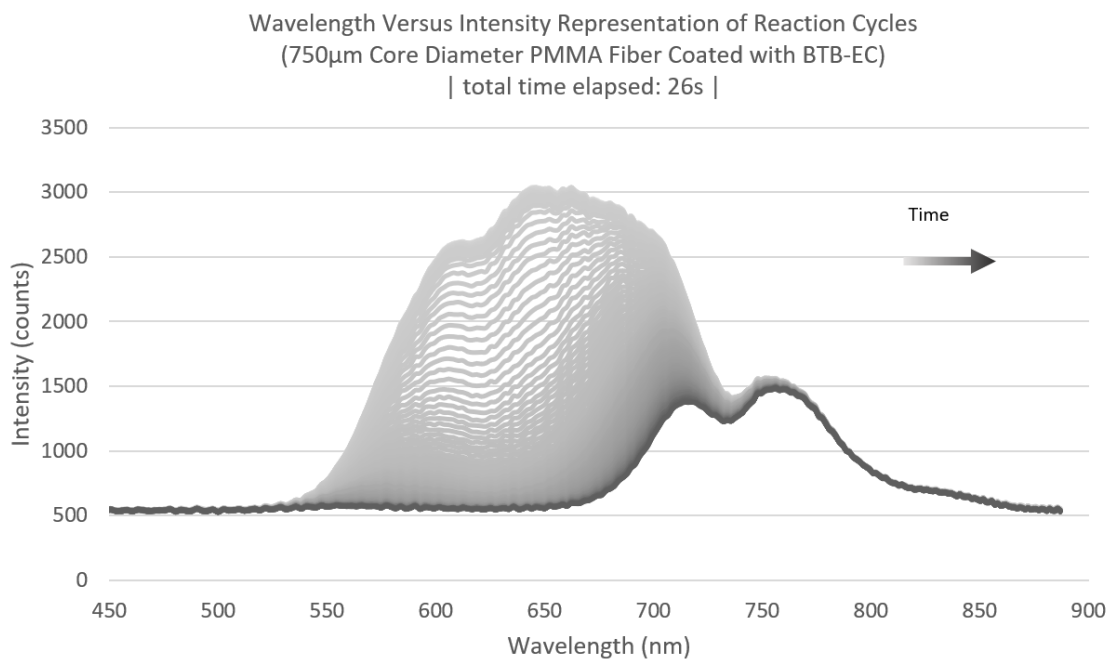

Figure 6: Transmission spectra through the FOCS, changing as the sensor is exposed to trace $NH_3$ (darker spectral lines correspond to later times). As expected, the final state has an intensity peak throughout the 700-800nm region.

Upon exposing a $750\mu m$ core diameter PMMA fiber coated in BTB-EC to trace $NH_3$, we typically observe a visible coloration response of the dye on the order of seconds, and a complete shift in evanescent field transmission spectra on the order of tens of seconds (fig. 6). Given the halogen light source and its respective intensity peaks, as expected, the transmission spectra shifts from an initial peak throughout ~500-800nm to ~700-800nm. This completely agrees with the previous analytic analysis — a confirmation of its reducing an otherwise complex full-wave analysis to a very good approximation, as permitted by the weakly guiding regime in the bessel and modified bessel modes.

At the time of writing, we have confirmed a minimum detectable concentration of 1ppm [25]. Of the fiber-optic chemical sensors that are readily scalable, portable, compatible to textile fibers in fabrics, and potentially wearable, this stands out as the best achieved so far. We expect this sensitivity to increase with further iterative cycles of testing and dye and core layer structural optimization guided by the analytics of the weak-guiding approximation.

The remaining characteristics that characterize this FOCS are selectivity and reversibility. FOCS are highly reversible as a function of analyte concentration exposure. When a detection event corresponds to small concentrations (~1-500ppm), regeneration is usually on the order of a few seconds to minutes (fig. 7). We found that a different sensing dye solution, composed of bromophenol blue (BPB) and EC (BPB-EC) regenerates significantly quicker (sometimes $1/10^{th}$ of the total regeneration time of BTB-EC). Exposure to much larger concentrations over prolonged periods could extend the regeneration time indefinitely, and, in extreme cases, even render the FOCS out of service. As for selectivity, we oserved a set of amines chemically similar to NH$_3$ that behave as interferents; otherwise, the FOCS remains highly selective. Of the interferents identified, we tested the effects of DEET and Spray 9, and observed much stronger selectivity to NH$_3$ vapor (fig. 8). One can imagine claiming a detection event only when the response is as strong as the target analyte, e.g., NH$_3$ in this case. Comparatively, however, this type of FOCS is highly selective and reversible.

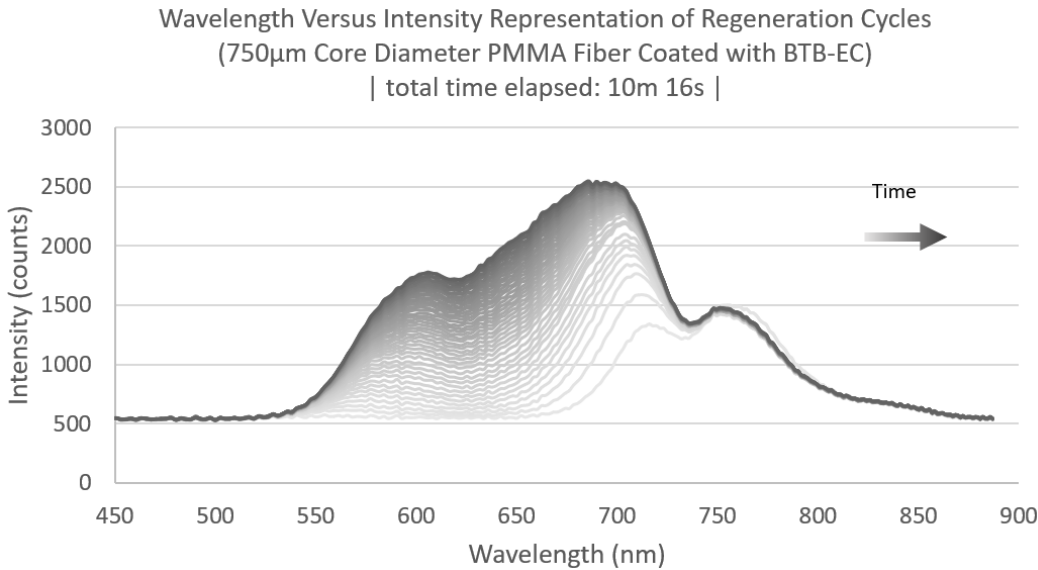

Figure 7: Transmission spectra through the FOCS, changing as the sensor regenerates from previous exposure to trace NH$_3$ (darker spectral lines correspond to later times). The FOCS was exposed to NH$_3$ vapor for 30s before it was purged with ambient air. The recovery process took a total of ~10 minutes for near complete regeneration.

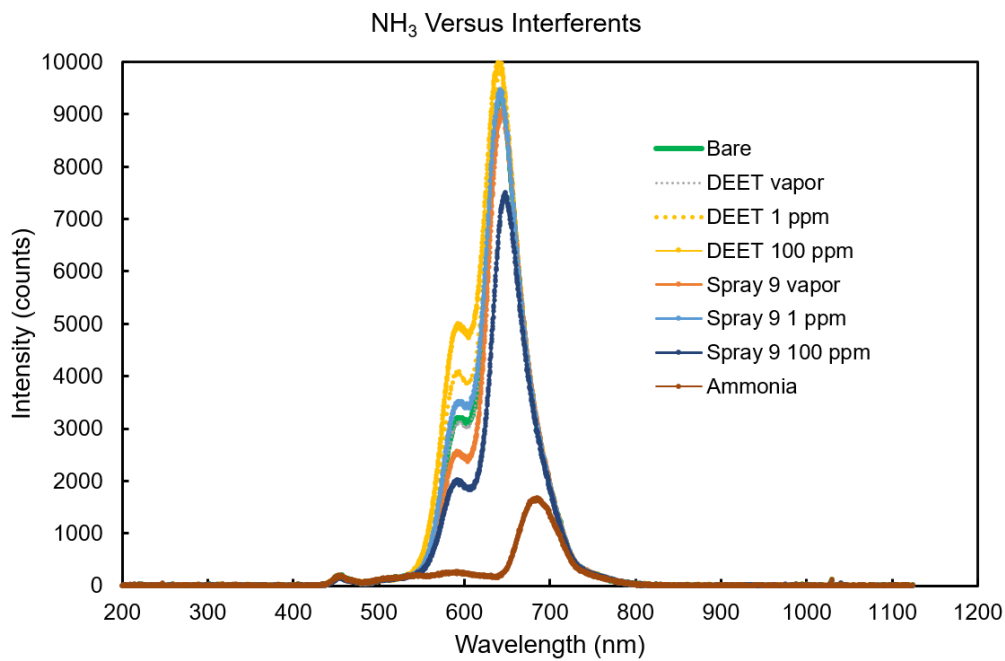

Figure 8: The effect of DEET and Spray 9 as interferents to the FOCS. We observe some response, but it is not nearly as strong as the NH$_3$ indicating relatively good selectivity even among interferents.

## Discussion and conclusion

The analytic formulation of the behavior of light in cylindrical dielectric multi-mode waveguides led to both a clearer understanding of FOCSs and a guide for further development and investigation [2][4][11][16][17][37][38]. Single-mode fibers are analytically tractable, but the need for a robust and multi-target FOCS requires multi-mode fibers. A full-wave analysis of multi-mode ($\sim 10^5$ modes) fibers is significantly more demanding and often obscures key features. Our analysis is much simpler, and captures the principal spectral responses.

The match between analytical results and experimental results were satisfactory. We showed the efficacy, robustness, and reversibility of FOCSs (among other desirable characteristics), placing them into a firm position for applications in many settings such as that of defense, commercial, industrial, and residential – corresponding to exemplary target analytes such as CWAs, TICs, EPs, etc., where we used $\text{NH}_3$ as a representative analyte. At the current reported sensitivity of 1ppm (and expectations of its increase with further optimization), and the affordability, ruggedness, selectivity, reversibility, portability, safety, and scalability of FOCS, this platform renders itself as a leading contender for (distributed) standoff sensing and monitoring of target chemicals. Of the ones scalable to hundreds meters and compatible to textile fibers and fabrics for portable, even wearable, applications, the 1ppm result makes it stand out already as the current best-in-class. Though selectivity is comparatively high, there is more work to be done on identifying and mitigating interferents. In passing, we mention that there are also some slight intolerances toward changes in humidity, which calls for more development and investigation. Otherwise, FOCSs seem to be a good option as a standoff and constant monitor for target analytes with strong applications to textiles and wide-area distributed sensor networks.

In sum, the analytic formulation of waveguide theory provided good supporting models and terminological foundation leading to the analytic formulation of multi-mode optical fibers, and here specifically, FOCSs. In a good correspondance between theory and experiment, the theoretical predictions of the propagation of light through an optical fiber were consistent with our experimental results.

We acknowledge sponsorship and funding from the National Science Foundation (CMMI-1530547), Defense Threat Reduction Agency, Air Force Office of Scientific Research (FA9550-19-1-0355), US Army Combat Capabilities Development Command - Soldier Center (CA W911QY-17-2-0001), and the Army Research Office (W911NF-17-S-0002).